\documentclass[journal]{IEEEtran}
\ifCLASSINFOpdf
\else
\fi

\usepackage{graphicx}
\usepackage{amsmath}
\usepackage{amssymb}
\usepackage{array}
\usepackage{booktabs,tabularx}
\usepackage{placeins}
\usepackage{color}
\usepackage{url}
\usepackage[disable]{todonotes}

\DeclareMathOperator*{\argmin}{arg\,min}
\DeclareMathOperator*{\argmax}{arg\,max}

\newcolumntype{C}[1]{>{\centering\let\newline\\\arraybackslash\hspace{0pt}}m{#1}}

\begin{document}
%
\title{Automatic Stress Detection in Working Environments from Smartphones' Accelerometer Data: A First Step}
%
%
%

\author{Enrique~Garcia-Ceja,
        Venet Osmani
		and Oscar Mayora%
\thanks{E. Garcia-Ceja is a PhD student at Tecnol\'{o}gico de Monterrey, Monterrey, M\'{e}xico. e-mail: e.g.mx@ieee.org}

\thanks{V. Osmani and O. Mayora are with CREATE-NET International Research Centre, Trento, Italy. e-mail: venet.osmani@create-net.org oscar.mayora@create-net.org}
\thanks{\textit{Author's copy of the paper published in IEEE Journal of Biomedical and Health Informatics}}

\thanks{\textit{Enrique Garica-Ceja, Venet Osmani, Oscar Mayora "Automatic Stress Detection in Working Environments from Smartphones’ Accelerometer Data: A First Step" IEEE Journal of Biomedical and Health Informatics, DOI:10.1109/JBHI.2015.2446195, 2015}}}

\maketitle

\begin{abstract}
Increase in workload across many organisations and consequent increase in occupational stress is negatively affecting the health of the workforce. Measuring stress and other human psychological dynamics is difficult due to subjective nature of self-reporting and variability between and within individuals. With the advent of smartphones it is now possible to monitor diverse aspects of human behaviour, including objectively measured behaviour related to psychological state and consequently stress. We have used data from the smartphone's built-in accelerometer to detect behaviour that correlates with subjects’ stress levels. Accelerometer sensor was chosen because it raises fewer privacy concerns (in comparison to location, video or audio recording, for example) and because its low power consumption makes it suitable to be embedded in smaller wearable devices, such as fitness trackers. 30 subjects from two different organizations were provided with smartphones. The study lasted for 8 weeks and was conducted in real working environments, with no constraints whatsoever placed upon smartphone usage. The subjects reported their perceived stress levels three times during their working hours. Using combination of statistical models to classify self reported stress levels, we achieved a maximum overall accuracy of 71\% for user-specific models and an accuracy of 60\% for the use of similar-users models, relying solely on data from a single accelerometer.
\end{abstract}


%
\IEEEpeerreviewmaketitle

\section{Introduction}
%
%
%
 
\IEEEPARstart{T}{he} competitive nature of the world economy and the use of advanced information and communication technologies has changed the nature of workplace environments, ensuring increased connectivity and consequently reachability of workers even outside working hours. This has resulted in an increase of workload \cite{Frese2000}, which has become a common issue in many organisations, where employees experience psychological problems related to occupational stress. According to the Fourth European Working Conditions Survey (EWCS), work-related stress was reported by 22\% of workers from 27 Member states of the European Union \cite{eu27}. Furthermore, higher prevalence of stress has been reported in North America, where 55\% of population has reported increased workload having a significant impact on physical and mental health as described in APA Survey \cite{APA2012}.

Occupational stress has been proven to contribute to disease activation. Several research studies have found that stress at work is associated with cardiovascular diseases \cite{cardio}, musculoskeletal diseases \cite{muscul}, immunological problems \cite{imun}, and problems with mental health such as anxiety and depression disorders \cite{Thoits2013}. In regard to organizational well-being, a decline of physical and mental health of workers has been reported in Paoli et al.\cite{Paoli2003}, leading to a decrease in the performance, decrease in overall productivity of organization and increased cost in terms of absenteeism. Experiencing work-related stress is common in working environments and low levels of stress can even result in productivity increase \cite{hillier2005wellness}. However, stress responses of employees are triggered when work-related pressure (such as quantity of work to be accomplished in a short period of time, pressure to work overtime, low social support, job insecurity and lesser breaks or holidays) challenge the human ability to cope with them. 

Considering detrimental effects of prolonged exposure to stress both for employees and organizations, there is a clear need for a system that can continuously monitor behaviour of workers and correlate various behaviour aspects with perceived stress levels. Several research works have used different sensing technologies, such as sound analysis \cite{he2009stress}, image processing from cameras \cite{valstar2011first} and physiological sensors \cite{demerouti2008oldenburg} to detect stress. Considering privacy concerns when using cameras and microphones, physiological measures have become an increasingly popular approach for measuring stress-related signs from sensor data (typically GSR and heart-rate sensors), such as work in \cite{Kocielnik2012}. However, there are several concerns about using physiological sensors, principally due to their obtrusive nature, lack of comfort and ability to be worn continuously \cite{Ikehara2005}, consequently impacting natural behaviour of the subjects.

With these points in mind, and based on our previous studies \cite{grunerbl2015, Gruenerbl2014}, smartphones have a distinct advantage in that they are already familiar and widely adopted devices, thus minimising "observer effect" and do not pose additional discomfort on the monitored subjects \cite{sano2013, bauer2012}. Using smartphones to monitor behaviour of subjects, we report the results of our study in detecting stress levels in real working environments. We recruited 30 subjects from two different organizations that participated in our 8 week study, where each participant reported perceived stress levels three times during working hours using self-assessment questionnaire.

Through the use of a combination of statistical models to classify self reported stress levels, we achieved an overall accuracy of 71\% for \emph{user-specific} models and 60\% for the use of \emph{similar-users} models. These results are comparable to the state of the art results in stress recognition, with the difference that our work relies solely on a single triaxial accelerometer sensor.
Furthermore, we also developed classification models using data from similar users, when building individual models was not feasible due to scarce data. Lastly, we evaluate the use of an ordinal classifier to take into account the class ordering information of the different stress levels.

Relying on a single accelerometer as the only sensor in detecting stress is especially promising when considering exponential rise of personal activity trackers (such as FitBit or Jawbone) that typically contain a single embedded accelerometer.

The rest of the paper is organized as follows: Section~\ref{sec:related_work} summarises previous research works for monitoring stress events from individuals in work- and real-life settings. 
Section~\ref{sec:collection} provides information about the group selection for the study, and how the data was collected. The details of data preprocessing are given in Section~\ref{sec:preprocessing}. Section~\ref{sec:eda} presents an exploratory data analysis as a first step towards building statistical classification models. Section~\ref{sec:models} presents the details of the different schemes used to classify stress levels. Section~\ref{sec:experiments} describes the experiments and results of our study. Conclusions and future research directions are given in Section~\ref{sec:conclusions}.

\section{Related Work}
\label{sec:related_work}

There have been several works that aim to detect stress in an automatic manner. For example, Carneiro et al. \cite{carneiro2012} used video cameras, accelerometers, touchscreens to extract different features while inducing different levels of stress during an electronic game session. Their experiments included 19 subjects and they used a J48 tree to classify touches as stressed or not achieving an accuracy of 78\%. In \cite{giakoumis2012}, Giakoumis et al. used video, accelerometers at the user's knees, galvanic skin response and electrocardiogram sensors to detect stress. There were 21 participants in their study and the Stroop color test \cite{Jensen1966} was used to induce stress. Their results showed that using behavioural features together with physiological measures helped to increase the stress detection accuracy compared when using just physiological features. Recently, there has also been research to detect stress outside lab environments by using wearable sensors. Lu et al. \cite{lu2012} implemented an application running in a smartphone to detect stress using voice as input. Sano \& Picard \cite{sano2013} used data collected from a wrist sensor, surveys and a mobile phone to classify stressed and not stressed states achieving results of over 75\% accuracy.

Two types of setups that have been used in previous works can be identified: \emph{In-lab experiments} and \emph{unconstrained experiments}. In-lab experiments are performed with controlled conditions, i.e., subjects are required to stay within an specific physical place and to follow a standard protocol. This protocol generally consists of filling surveys and performing a series of experiments in a specific order. 
In an unconstrained setup, the subject is generally given a set of wearable sensors and the data is collected while the subject performs their daily routines without following any predefined schedule.


Table \ref{tab:classification} presents a summary of related works on automatic stress detection and classified according to the type of experiment: \emph{In-lab}, \emph{unconstrained} and the type of stressors: \emph{controlled}, \emph{uncontrolled} and \emph{unknown}. This work differs from the previous work in the following aspects: 1) The data was collected in an unconstrained out of the lab environment and with unknown stressors using only an accelerometer sensor from smartphone; 2) We explore the potential of using data from a single source (accelerometer) to detect acute stress levels. We chose this sensor because it is non-visual and non-auditory, and thus mitigates privacy concerns and does not interfere with the individual's daily routines  \cite{matic2012multi, matic2013, matic2012analysis}; and 3) We built classification models using data from similar users in cases when building individual models is not feasible due to scarce data.

\begin{table}[t!]
  \centering
  \caption{Classification of different related works. Type column indicates if the experiment was performed in-lab or in an unconstrained environment and the type of stressors used: controlled, uncontrolled, unknown.}
  \label{tab:classification}
    \begin{tabular}{C{1.3cm}C{1.3cm}C{2.0cm}C{2.3cm}}
    \hline
    \textbf{Work} & \textbf{Type} & \textbf{Data sources} & \textbf{Details} \\
    \hline
    Carneiro et al. \cite{carneiro2012} & In-lab, controlled & video cameras, accelerometers, pressure-sensitive touchscreens & 19 subjects. 78\% accuracy in classifying touches as stressed or not using a J48 tree. \\
	\hline
	Giakoumis et al. \cite{giakoumis2012} & In-lab, controlled & video, accelerometers at users' knees, Galvanic skin response, electrocardiogram & 21 subjects. Avg. accuracy of 100\% for their dataset 1 and 96.6\% for dataset 2 when using all sensors. \\
    \hline
    Sun et al. \cite{sun2012} & In-lab, controlled & electrocardiogram, galvanic skin response, accelerometer & 20 subjects. Overall accuracy 92.4\% for 10-fold cross validation and 80.9\% between subjects classification. \\
    \hline
    Bauer \& Lukowicz \cite{bauer2012} & unconstrained, uncontrolled & gps, wi-fi, bluetooth, call logs, sms & 7 subjects. Detected a change of behaviour during stress periods of approx. 86\% of the participants. \\
    \hline
    Lu et al. \cite{lu2012} & In-lab, unconstrained, uncontrolled & audio & 14 subjects. accuracy of 81\% and 76\% for indoor and outdoor environments with model adaptation. \\
    \hline
    Muaremi et al. \cite{muaremi2013} & unconstrained,  unknown & heart rate, audio, acceleration, gps, calls, contacts, etc. & 35 subjects. 61\% accuracy for user specific models and 53\% for a general model. \\
    \hline
    Sano \& Picard \cite{sano2013} & unconstrained,  unknown & accelerometer, skin conductance, calls, sms, location, screen & 18 subjects. Accuracies over 75\% \\
    \hline
    Bogomolov et al. \cite{bogomolov2014} & unconstrained, unknown & call logs, sms, bluetooth & 117 subjects. An overall recognition accuracy of 72.39\% with Random forest model. \\
    \bottomrule
    \end{tabular}
\end{table}

\section{Data Collection}
\label{sec:collection}

Behavioural data were obtained using the built-in sensors of Samsung Galaxy SIII Mini smartphones. The data were collected with the written, informed consent of all participants and stored in the memory of the smartphone using the application developed by our team. Additional information pertaining to the usage of apps and contextual information such as location, accelerometer, social activities, phone calls, SMS, Wi-Fi, and proximity was also recorded. However, in this work we analysed only the data from the triaxial accelerometer, recorded continuously. Given that the phone application collected data from several sensors, the accelerometer sampling rate was set at 5 Hz in order to optimise the battery life. This was adequate for our analysis since work in \cite{szhang2012} showed that with a sampling rate of 5 Hz it is still possible to recognize physical activities with an accuracy of 94.98\%, while we are not analysing short, fine grained movements (as in activity or gesture recognition) but rather focus on the overall behaviour that spans several minutes.

We also collected subjective information related to subjects' stress and psychological states involving a series of questions/answers gathered from a survey. These questions were derived from a clinically validated burnout questionnaire; the Oldenburg Burnout Inventory (OLBI) \cite{demerouti2008oldenburg}.
Subjective psychological scores for stress, were reported in a questionnaire three times during the working days (morning, afternoon, and end of workday) on a 5-point scale. This information was then converted into an ordinal scale to represent stress levels as \emph{low, medium} and \emph{high}, due to inherent differences in subjective reporting of stress levels between individuals and also within individuals \cite{ClusterBasedStress2015}, that is, for one user the value of 4 may mean 'highly stressed' whereas for another 'a little bit above normal'. Grouping the ratings into a smaller number of ordinal points alleviates some of the inherent subjectivity.
\subsection{Participants}
Sensor data was collected from 30 healthy subjects and analysed with self-reported stress data for a period of 8 weeks, excluding weekends. Due to user compliance issues, the average number of data collection days per user was $29\pm 6$. Furthermore, some surveys during the day were occasionally skipped by the users. The participants used the phone from morning until the end of the work day, without any restrictions whatsoever placed upon the use of the phone in a specific manner.
In order to get insights in the working style and gain more knowledge from employees in their working environments, we chose to recruit participants from two different companies located in province of Trentino, Italy. The study involved 18 (60\%) males and  12 (40\%) females aged 37.46$\pm$7.26 years. Participants were informed that the goal of the study was to monitor behaviour activities relevant to stress. All participants consented to participate in the study and to have their data recorded. They were also informed that all the collected data was anonymous and will be used for research purposes only.

\section{Pre-processing}
\label{sec:preprocessing}

\emph{Feature extraction}

From the raw accelerometer data a total of 34 features from time and frequency domain were extracted. The feature extraction was performed on non-overlapping fixed length windows of 128 samples (25.6 seconds.). The 34 features were: \emph{Mean x axis}, \emph{Mean y axis}, \emph{Mean z axis}, \emph{StdDev x axis}, \emph{StdDev y axis}, \emph{StdDev z axis}, \emph{Variance x axis}, \emph{Variance y axis}, \emph{Variance z axis}, \emph{Variance 3 axes}, \emph{Mean 3 axes}, \emph{Max 3 axes}, \emph{Min 3 axes}, \emph{Standard Deviation 3 axes}, \emph{Absolute Value 3 axes}, \emph{Median 3 axes}, \emph{Range 3 axes}, \emph{Variance Sum} \cite{funf}, \emph{Magnitude} Eq.(\ref{eq:magnitude}), \emph{Signal Magnitude Area} Eq.(\ref{eq:sma}), \emph{Root Mean Squared} Eq.(\ref{eq:rms}), \emph{Curve Length} Eq.(\ref{eq:cl}), \emph{Non Linear Energy} \cite{mukhopadhyay1998new}, \emph{Entropy:} differential entropy from time domain magnitude Eq.(\ref{eq:de}) \cite{cover2012elements}, \emph{Energy:} which is the sum of the squared discrete FFT component magnitudes of the signal. Eq.(\ref{eq:energy}) \cite{bao2004}, \emph{Mean Energy}, \emph{StdDev Energy}, \emph{DFT (Discrete Fourier Transform)}, \emph{Peak Magnitude} which is the maximum value of the magnitude. Eq.(\ref{eq:pm}), \emph{Peak Magnitude Frequency} which is the frequency that corresponds to the maximum magnitude. Eq.(\ref{eq:pmf}), \emph{Peak Power} which is analogous to peak magnitude but on the power spectrum, \emph{Peak Power Frequency} this is analogous to peak magnitude frequency, \emph{Magnitude Entropy} Eq.(\ref{eq:me}) and \emph{Power Shannon Entropy} same as Magnitude Entropy but over the power spectrum.

\begin{equation}
Magnitude = \frac{1}{n}\sum\limits_{i = 1}^n {\sqrt {x_i^2 + y_i^2 + z_i^2} }
\label{eq:magnitude}
\end{equation}

\begin{equation}
SMA=\frac{1}{T}\int_o^T{x(t)|dt}+\int_o^T{y(t)|dt}+\int_o^T{z(t)|dt}
\label{eq:sma}
\end{equation}

\begin{equation}
RMS = \sqrt {\frac{1}{n}\left( {x_1^2 + x_2^2 +  \cdots  + x_n^2} \right)}
\label{eq:rms}
\end{equation}

\begin{equation}
curve length = \sum\limits_{i = 2}^N {\left| {{x_{i - 1}} - {x_i}} \right|}
\label{eq:cl}
\end{equation}

\begin{equation}
h(X) = \int_\mathbb{X} {f(x)\log f(x)dx}
\label{eq:de}
\end{equation}

\begin{equation}
energy = \sum\limits_{i = 1}^{(n/2)} {x{{[i]}^2}}
\label{eq:energy}
\end{equation}

\begin{equation}
pm = \max_{i=1..(n/2)}x_i
\label{eq:pm}
\end{equation}

\begin{equation}
pmf = \argmax_{i=1..(n/2)}x_i
\label{eq:pmf}
\end{equation}

\begin{equation}
H(X) =  - \sum\limits_{i = 0}^{N - 1} {{p_i}{{\log }_2}{p_i}}
\label{eq:me}
\end{equation}

\emph{Self-reported Stress} 

The stress scale in the survey has the scale 1 to 5, where 1 means least stressed and 5 means most stressed. For the purpose of our analysis we grouped those values into three groups: \emph{low stress} for values of 1 and 2; \emph{medium stress} for a value of 3; and \emph{high stress} for values of 4 and 5.

For our analysis, we considered observations from the second and third surveys only because there is no accelerometer data before the first survey (beginning of the day). To characterize each survey, we took the features from the previous 2 hours for each survey and computed summary statistics which will be used as the final features: mean, maximum and minimum value of each of the 34 features giving a total of 102 features. Table \ref{tab:total_observations} shows the total number of observations for each of the stress levels.

\begin{table}[h]
  \centering
  \caption{Total number of observations for the second and third surveys}
    \label{tab:total_observations}
    \begin{tabular}{ccccc}
    Stress level: & \textbf{Low} & \textbf{Medium} & \textbf{High} & \\
    \midrule
     \# observations & 667 & \multicolumn{1}{c}{521} & \multicolumn{1}{c}{329} & Total: 1,517 \\
    \bottomrule
    \end{tabular}
\end{table}

\section{Exploratory Data Analysis}
\label{sec:eda}

\begin{figure}[h]
\centering
\includegraphics[scale=0.44]{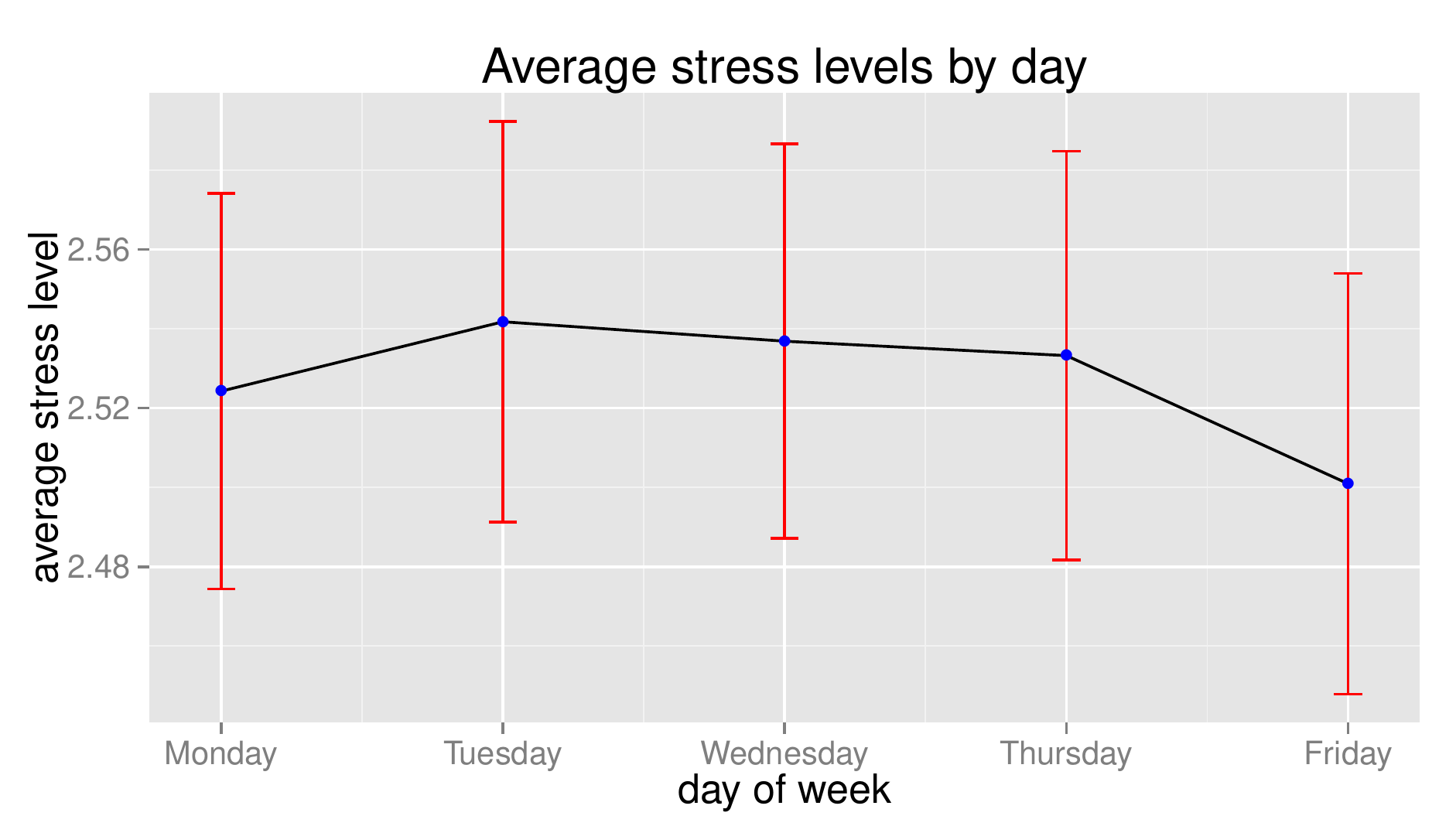}
\caption{Average stress levels by weekday with standard error bars of the mean.}
\label{fig:stress_by_day}
\end{figure}

In this section we present a general overview of the data. Figure \ref{fig:stress_by_day} shows the average of the self reported stress level scores by weekday over all users using data from the 3 surveys. It can be seen that the maximum stress level is reported on Tuesday and then begins to decrease towards its minimum on Friday. The resulting standard error bars overlap with each other suggesting that the differences between days are not significant, confirmed with an analysis of variance test.

\begin{figure}[h]
\centering
\includegraphics[scale=0.63]{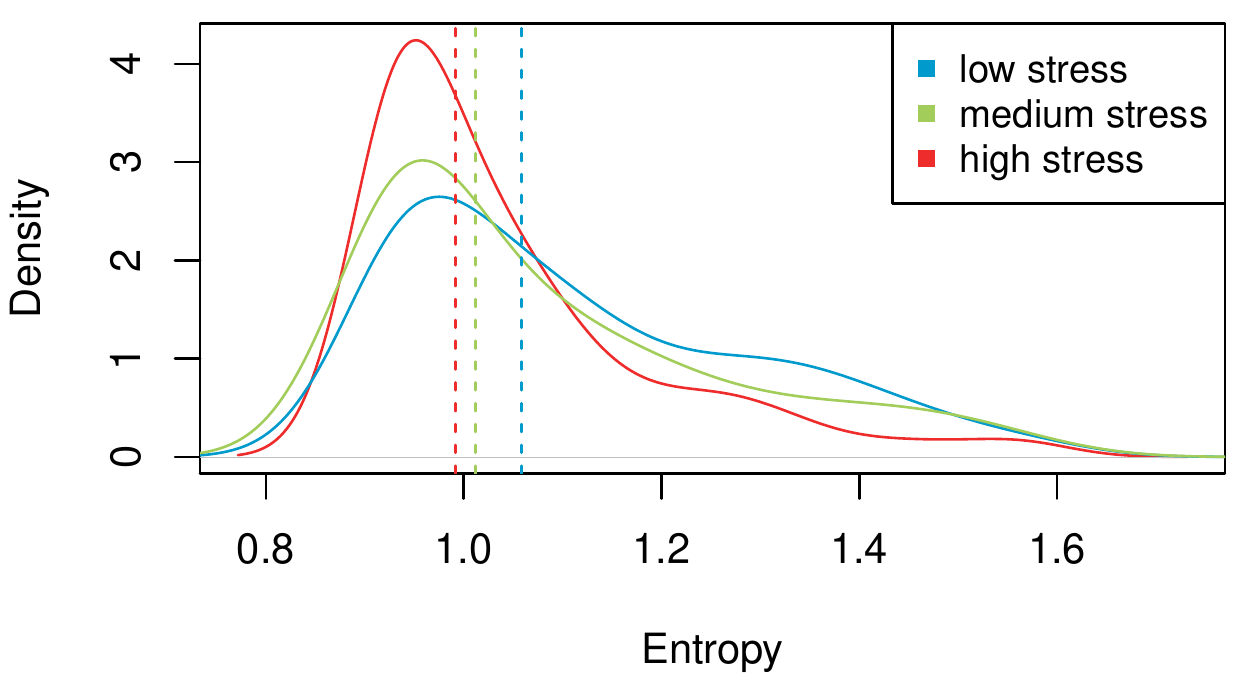}
\caption{Estimated density for Entropy feature. Vertical lines represent the median}
\label{fig:density_entropy}
\end{figure}

Now we investigate whether extracted features could be used as potential predictors for stress levels. Figure \ref{fig:density_entropy} shows the estimated density function for the \emph{Entropy} feature over all users. Vertical lines indicate the median. Through visual inspection, the difference between the median of the Entropy for high stress is clearly visible from that of low stress. The difference between medium and low stress is also clear and the difference between high and medium is still noticeable but smaller. It seems that Entropy is a good candidate feature (independently of the others) to differentiate between high/low and medium/low stress levels but it may have difficulties differentiating between high/medium stress levels.

To see whether or not a specific feature is significantly different between every pair of possible stress levels (low/high, low/medium and high/medium) for each of the users, a Mann Whitney U test \cite{sheskin2003} was performed with a significance level $\alpha=0.01$ and bonferroni p-values correction. This test was chosen because it is non-parametric and most of the feature distributions are not normal. The results of the statistical test indicated that for most of the features and users the differences were significant (except for the Peak Magnitude feature). However, this does not necessarily mean that most of the features will be good predictors since the differences may be to small to be detected or to be useful to a given classifier model. In order to check the effect size of each of the features we computed the Cohen's d effect size and quantified it using the thresholds defined in \cite{cohen1992}, i.e., $|d| < 0.2$ `negligible', $|d| < 0.5$ `small', $|d| < 0.8$ `medium', otherwise `large'. The results of this test indicated that for almost half of the features the effect size was at least medium. Despite the fact that almost all features are different for each of the stress levels, their effect sizes are small and just a few of them are medium or large for some of the users (details of the statistical results for each feature were omitted due to space constraints). These exploratory results suggest that some of the features (independently of the others) can be used as potential predictors of stress levels. Next, we will use multivariate statistical models and  feature selection to find combinations of good discriminative features to detect stress levels.  Since we quantified the original stress levels as three different classes \{low, medium, high\} we will state the problem as a classification problem. Given the set of computed features from the accelerometer data we want to predict the users' self-reported stress levels. In this case we will use multivariate classification models which are discussed in the next section.

\section{Statistical Models}
\label{sec:models}

The results from the exploratory data analysis suggest that we could use some of the features as predictors to classify the different stress levels. For this purpose, we are going to use two classification models namely: 1) Naive Bayes \cite{Witten2011} (pp. 90-97) and 2) Decision Trees \cite{brian1996} (ch. 7).

As we discussed earlier, some of the features may increase the performance of the classifiers while others may have the opposite effect. To find good combinations of features to build the models we used a feature selection method called Forward Feature Selection \cite{james2014} (pp. 207) which consists of adding predictors one by one to the model and at each step the variable that increases performance criteria the most is retained. In this case we used accuracy as the performance criteria.

\subsection{Model Schemes}
\label{sec:schemes}

In recent related works it has been common to build \emph{user-specific} and \emph{general} models to classify stress levels \cite{muaremi2013,lu2012}. For the \emph{user-specific} case, individual models are trained and evaluated for each of the users using their own data. The \emph{general model} consists of building the model with data from all the users. This can be done by aggregating all the data from all the users or for each specific user $i$ build a model with the data from all other users $j$, $j \neq i$ and test the model with the data from user $i$. The latter approach is sometimes referred to as \emph{leave one person out}. In Lu et al. \cite{lu2012} they also used an hybrid approach called  model adaptation which starts with a general model and gets adapted to each individual as more data is available.

Following this methodology, we used the \emph{user-specific} and the \emph{general model} approach. For the \emph{general model} we used the \emph{leave one person out} scheme. We also built \emph{similar-users models} which differ from the general model in that instead of building one model for a user $i$ using observations from all other users $j$,$j \neq i$, the model for user $i$ is built using observations from just a subset of similar users. The rationale behind this scheme is that for any two users, their behavioural patterns across stress levels may be different. For example, a user may tend to be more active when he is stressed but another one may tend to be more sedentary when stressed.

Building a single model that includes users with different behaviour patterns is not desirable since this will introduce noise. Rather, we may want to build a model for a specific user with data just from similar users. In this case, even if there is not yet enough data to build an \emph{user-specific} model a system could build a model from similar users and start giving feedback until there is sufficient data to build an individual model.


\emph{Similar-users Model:} Here, the idea is to build a model to predict stress levels for the test user $u_t$ using data from the set of users $\mathbb{S}$, where $\mathbb{S}$ is the set of users with similar behaviour to $u_t$. The behaviour of each user will be represented by a single vector $\textbf{b}_i$ of size = $\binom {|C|} {2}|F|$ where $|C|$ is the number of classes and $|F|$ is the number of features. In this case $\binom {|C|} {2} = 3$ which corresponds to every possible combination of stress levels: \emph{low-medium, low-high, medium-high}. For each feature we want to know how does the median value changes between the different pairs of stress levels. For example, for one user the difference between $median(f_{1_{low}}) - median(f_{1_{high}})$ may be positive but for other user it may be negative where $median(f_{1_{low}})$ is the median of a specific feature when the stress level is low (and the same applies for all other levels). The behaviour vector $\textbf{b}_i$ is constructed by computing for each feature, the difference of the medians between every pair of stress levels.

To find $\mathbb{S}$ we used k-means clustering to group the behaviour vectors $\textbf{b}_i$, $i \neq t$ into $k$ groups $G_{1..k}$ and let $\mathbb{S}$ be the group who's centroid has the minimum distance to the behaviour vector of the test user $\textbf{b}_t$, i.e., $\mathbb{S}=\argmin_{G_{1..k}} dist(\textbf{b}_t,centroid(G_i))$. Since $u_t$ is the test user, $\textbf{b}_t$ is computed using only a random subset $O_{t,p}$ of the total observations of user $t$ where $p$ indicates what percentage of the total observations was taken. The subset $O_{t,p}$ of observations that was used to construct $\textbf{b}_t$ to find the similar users is discarded when evaluating the model to avoid over-fitting.

\begin{figure}[!h]
\centering
\includegraphics[scale=0.8]{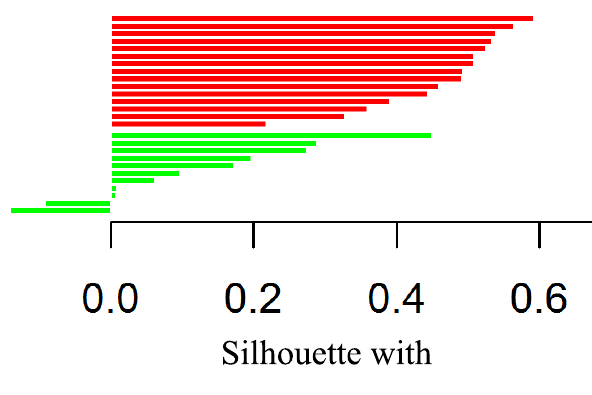}
\caption{Silhouette plot for $k=2$ with resulting silhouette index of 0.32. Line colors represent the different clusters.}
\label{fig:sil_k2}
\end{figure}

\begin{figure}[!h]
\centering
\includegraphics[scale=0.8]{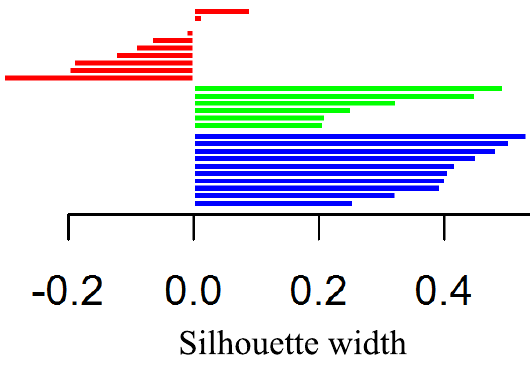}
\caption{Silhouette plot for $k=3$ with resulting silhouette index of 0.2. Line colors represent the different clusters.}
\label{fig:sil_k3}
\end{figure}

The k-means algorithm requires to specify the number $k$ of desired groups. To find a good approximation of $k$ we used the \emph{silhouette} index \cite{rousseeuw1987} which is a measure of the quality of the resulting groups. The k-means algorithm is run for $k=2,3,..,upper bound$ and the $k$ that maximizes the silhouette index is chosen as the final number of groups. Figure \ref{fig:sil_k2} shows an example of the resulting silhouette plot when grouping similar users to build a model for some specific subject when $k=2$. In this plot each line represents a behaviour vector $\textbf{b}$ and its length represents its silhouette width $s(i)$. A $s(i)$ close to 1 means that the feature vector $i$ is well clustered, i.e., there is little doubt that $i$ has been assigned to an appropriate group. The overall silhouette index is the average of all $s(i)$ and in this case it was 0.32. It can be seen that some feature vectors had a silhouette width less than or close to 0. This means that it is not clear whether these feature vectors should have been assigned to another cluster. Figure \ref{fig:sil_k3} shows the silhouette plot for $k=3$. In this case the silhouette index was 0.2 which is much lower and in the first cluster almost all data points have a silhouette width close to 0. For $k=4,5$ the silhouette index was 0.2 and 0.18 respectively, thus, $k=2$ was chosen as the number of final clusters for this specific user. Note that the plots have 26 bars (users) instead of the expected 29. This is because some users did not report \emph{high} stress levels and thus they have missing values in their behaviour vectors in which case they were excluded from the clustering phase. On the other hand, if the test user $u_t$ did not report \emph{high} stress levels, the columns with \emph{high} stress levels of the other users' behaviour vectors are truncated and thus, all other 29 users were included in the clustering procedure.

\subsection{Ordinal Classification}
\label{sec:ordinal}

Typically, classification algorithms assume that the response class is unordered but there are situations in which there is a natural ordering of the response variable, i.e., an ordinal class. Ordinal variables are typically found in surveys' responses for example, \emph{Very poor, Poor, Fair, Good, Excellent}. For our case we have: \emph{low} $<$ \emph{medium} $<$ \emph{high} stress levels. In order to take into account this  ordering information, we also implemented an ordinal classification approach described by Frank \& Hall \cite{Frank2001} which enables standard classification algorithms to make use of ordering information. This approach consists of transforming a k-class ordinal problem into k-1 binary class problems and computing the probability of each of the k ordinal classes. The final prediction is the class with maximum probability. We applied this approach with the Naive Bayes classifier. 

From the performance measurement point of view, usually, the classifiers are assessed with measures appropriate for unordered classes. These measures treat all errors as equal, e.g., confusing \emph{low} with \emph{medium} has the same error weight as confusing \emph{low} with \emph{high} but clearly, the latter error should be more severely penalized as discussed in \cite{muaremi2013}. In \cite{gaudette2009} several performance measures for ordinal classes were evaluated. For example, \emph{Mean Squared Error} (MSE) is more suitable when the severity of the errors is more important while \emph{Mean Absolute Error} (MAE) is preferred in situations where the tolerance for small errors is lower. Another performance measure is the \emph{Linear Correlation}. A strong correlation between the predictions and the ground truth is an indication of a good classifier. A more optimistic measure is the \emph{Accuracy within n} (ACC1, ACC2,.., ACC\emph{n}) which allows a wider range of outputs to be considered correct. For example, if the correct output is 4, outputs of 3,4 and 5 are considered as correct for $n=1$, i.e., ACC1. The usual accuracy measure would be ACC0.

\section{Experiments and Results}
\label{sec:experiments}

In this section we present the results for the three model schemes discussed in Section \ref{sec:schemes}: \emph{user-specific}, \emph{general} and \emph{similar-users} models. In the Feature Forward Selection step, for each of the candidate feature subsets 5-fold cross validation is performed in the case of \emph{user-specific} models and \emph{leave one person out} cross validation for the \emph{general} and \emph{similar-users} models. For the \emph{similar-users} model, 50\% of the data was used to find the most similar users, i.e., $O_{i,50}$. We used the following performance measures that take into account the ordinal nature of the response variable to evaluate the models: Mean Absolute Error (MAE), Root Mean Squared Error (RMSE), Pearson Correlation (Pearson cor), Spearman Correlation (Spearman cor) and Accuracy within 1 (ACC1).

In our experiments, 4 classifiers were used: \emph{Naive Bayes}; \emph{Decision Tree}; \emph{Ordinal Naive Bayes} which uses the approach described in Section \ref{sec:ordinal}; and as a baseline a \emph{Random} classifier which randomly predicts a class based on their prior probabilities. Table \ref{tab:results_individual} shows the results for the \emph{user-specific} models. Here we can see that all classifiers (except Random) had a similar overall performance. Note that the ACC1 measure is very optimistic. The Random classifier had an ACC1 = 0.81. this is because an output will be counted as an error only if the  prediction is \emph{low} and the actual class is \emph{high} or vice versa. For the \emph{user-specific} case, the 10 most frequently selected features (in descending order) were: Magnitude, Standard Deviation of the 3 axes, Minimum Energy, Maximum of the 3 axes, Peak Magnitude Frequency, Minimum variance Y, Maximum of variance sum, Max Range of the 3 axes, Maximum Mean Energy and Variance sum.

\begin{table*}[t]
\centering
\caption{User-specific model results}
\label{tab:results_individual}
\begin{tabular}{r|ccc|ccc|ccc|ccc|}
  \cline{2-13}
        & \multicolumn{3}{c|}{\textbf{Random}} & \multicolumn{3}{c|}{\textbf{Naive Bayes}} & \multicolumn{3}{c|}{\textbf{Ordinal Naive Bayes}} & \multicolumn{3}{c|}{\textbf{Decision Tree}} \\
  \cline{2-13}
        & \textbf{low} & \textbf{medium} & \textbf{high} & \textbf{low} & \textbf{medium} & \textbf{high} & \textbf{low} & \textbf{medium} & \textbf{high} & \textbf{low} & \textbf{medium} & \textbf{high} \\
  \cline{2-13}
  \textbf{Sensitivity} & 0.57     & 0.41     & 0.34     & 0.82     & 0.65     & 0.59    & 0.82     & 0.62     & 0.57    & 0.79     & 0.66     & 0.62 \\
  \textbf{Specificity} & 0.68     & 0.69     & 0.8     & 0.81     & 0.82     & 0.91    & 0.79     & 0.82     & 0.91    & 0.82     & 0.83     & 0.9 \\
  \textbf{Precision} & 0.58     & 0.41     & 0.32     & 0.77     & 0.65     & 0.66    & 0.75     & 0.65     & 0.65    & 0.77     & 0.67     & 0.63 \\
  \hline
  \textbf{Accuracy} & \multicolumn{3}{c|}{0.46} & \multicolumn{3}{c|}{0.71} & \multicolumn{3}{c|}{0.7} & \multicolumn{3}{c|}{0.71} \\
  \textbf{MAE} & \multicolumn{3}{c|}{0.66} & \multicolumn{3}{c|}{0.33} & \multicolumn{3}{c|}{0.35} & \multicolumn{3}{c|}{0.34} \\
  \textbf{RMSE} & \multicolumn{3}{c|}{0.96} & \multicolumn{3}{c|}{0.65} & \multicolumn{3}{c|}{0.67} & \multicolumn{3}{c|}{0.68} \\
  \textbf{Pearson cor} & \multicolumn{3}{c|}{0.24} & \multicolumn{3}{c|}{0.63} & \multicolumn{3}{c|}{0.62} & \multicolumn{3}{c|}{0.61} \\
  \textbf{Spearman cor} & \multicolumn{3}{c|}{0.25} & \multicolumn{3}{c|}{0.64} & \multicolumn{3}{c|}{0.62} & \multicolumn{3}{c|}{0.62} \\
  \textbf{ACC1} & \multicolumn{3}{c|}{0.86} & \multicolumn{3}{c|}{0.95} & \multicolumn{3}{c|}{0.95} & \multicolumn{3}{c|}{0.94} \\
  \cline{2-13}
\end{tabular}
\hfill
\end{table*}

Table \ref{tab:results_general} shows the results for the \emph{general} model scheme. As expected, the overall performance is much lower than that of the \emph{user-specific} scheme. Again all classifiers (except Random) had similar overall performances. The Random classifier had Pearson and Spearman correlations close to 0 while for the other models the correlation was stronger but still weak. The Ordinal Naive Bayes classifier did not present any improvement over the traditional Naive Bayes. The reason of this lack of improvement when including ordering information may be that in this case the number of classes is just 3 and as suggested by Frank \& Hall \cite{Frank2001} ``ordering information becomes more useful as the number of classes increases.''.

Table \ref{tab:results_hybrid} shows the results for the \emph{similar-users} model for Naive Bayes and Decision tree. The Ordinal Naive Bayes was omitted since it did not present any performance improvement in the previous cases. With respect to the general model, the \emph{similar-users} model had an increase of 8\% in accuracy for Naive Bayes and 5\% for the Decision Tree.

\begin{table*}[t]
\centering
\caption{General model results}
\label{tab:results_general}
\begin{tabular}{r|ccc|ccc|ccc|ccc|}
  \cline{2-13}
        & \multicolumn{3}{c|}{\textbf{Random}} & \multicolumn{3}{c|}{\textbf{Naive Bayes}} & \multicolumn{3}{c|}{\textbf{Ordinal Naive Bayes}} & \multicolumn{3}{c|}{\textbf{Decision Tree}} \\
  \cline{2-13}
        & \textbf{low} & \textbf{medium} & \textbf{high} & \textbf{low} & \textbf{medium} & \textbf{high} & \textbf{low} & \textbf{medium} & \textbf{high} & \textbf{low} & \textbf{medium} & \textbf{high} \\
  \cline{2-13}
  \textbf{Sensitivity} & 0.41     & 0.33     & 0.24     & 0.94     & 0.18     & 0.2    & 0.91     & 0.18     & 0.22    & 0.84     & 0.19     & 0.28 \\
  \textbf{Specificity} & 0.58     & 0.63     & 0.77     & 0.3     & 0.91     & 0.95    & 0.33     & 0.89     & 0.94    & 0.39     & 0.85     & 0.91 \\
  \textbf{Precision} & 0.43     & 0.32     & 0.23     & 0.51     & 0.53     & 0.55    & 0.51     & 0.47     & 0.54    & 0.52     & 0.41     & 0.47 \\
  \hline
  \textbf{Accuracy} & \multicolumn{3}{c|}{0.35} & \multicolumn{3}{c|}{0.52} & \multicolumn{3}{c|}{0.51} & \multicolumn{3}{c|}{0.5} \\
  \textbf{MAE} & \multicolumn{3}{c|}{0.83} & \multicolumn{3}{c|}{0.62} & \multicolumn{3}{c|}{0.61} & \multicolumn{3}{c|}{0.62} \\
  \textbf{RMSE} & \multicolumn{3}{c|}{1} & \multicolumn{3}{c|}{0.95} & \multicolumn{3}{c|}{0.94} & \multicolumn{3}{c|}{0.94} \\
  \textbf{Pearson cor} & \multicolumn{3}{c|}{0.01} & \multicolumn{3}{c|}{0.32} & \multicolumn{3}{c|}{0.33} & \multicolumn{3}{c|}{0.31} \\
  \textbf{Spearman cor} & \multicolumn{3}{c|}{0.01} & \multicolumn{3}{c|}{0.32} & \multicolumn{3}{c|}{0.33} & \multicolumn{3}{c|}{0.31} \\
  \textbf{ACC1} & \multicolumn{3}{c|}{0.81} & \multicolumn{3}{c|}{0.85} & \multicolumn{3}{c|}{0.86} & \multicolumn{3}{c|}{0.87} \\
  \cline{2-13}
\end{tabular}
\hfill
\end{table*}

\begin{table*}[t]
\centering
\caption{Similar-users model results}
\label{tab:results_hybrid}
\begin{tabular}{r|ccc|ccc|}
  \cline{2-7}
        & \multicolumn{3}{c|}{\textbf{Naive Bayes}} & \multicolumn{3}{c|}{\textbf{Decision Tree}} \\
  \cline{2-7}
        & \textbf{low} & \textbf{medium} & \textbf{high} & \textbf{low} & \textbf{medium} & \textbf{high} \\
  \cline{2-7}
  \textbf{Sensitivity} & 0.6     & 0.58     & 0.6     & 0.64     & 0.64     & 0.24 \\
  \textbf{Specificity} & 0.83     & 0.69     & 0.86     & 0.76     & 0.58     & 0.95 \\
  \textbf{Precision} & 0.73     & 0.5     & 0.55     & 0.67     & 0.44     & 0.59 \\
  \hline
  \textbf{Accuracy} & \multicolumn{3}{c|}{0.6} & \multicolumn{3}{c|}{0.55} \\
  \textbf{MAE} & \multicolumn{3}{c|}{0.45} & \multicolumn{3}{c|}{0.49} \\
  \textbf{RMSE} & \multicolumn{3}{c|}{0.75} & \multicolumn{3}{c|}{0.76} \\
  \textbf{Pearson cor} & \multicolumn{3}{c|}{0.52} & \multicolumn{3}{c|}{0.43} \\
  \textbf{Spearman cor} & \multicolumn{3}{c|}{0.52} & \multicolumn{3}{c|}{0.44} \\
  \textbf{ACC1} & \multicolumn{3}{c|}{0.94} & \multicolumn{3}{c|}{0.95} \\
  \cline{2-7}
\end{tabular}
\end{table*}

\begin{figure}[!h]
\centering
\includegraphics[scale=0.65]{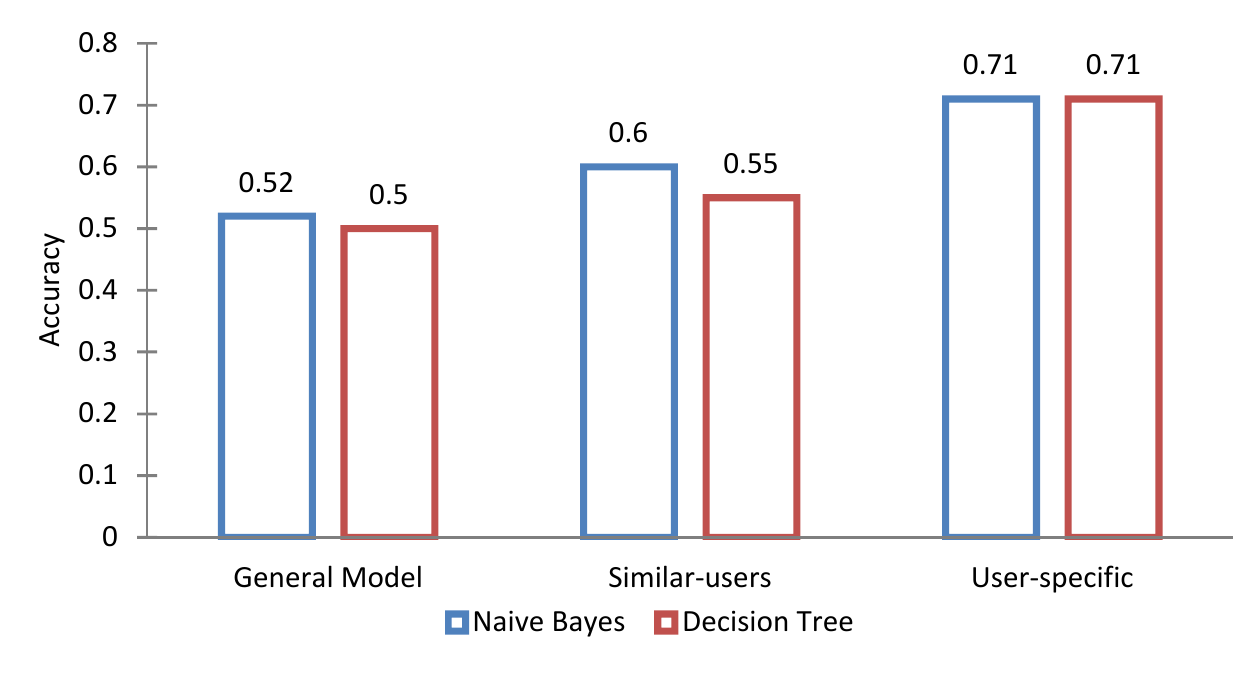}
\caption{Comparison between general models, similar users models and user specific models.}
\label{fig:barplot}
\end{figure}

\section{Conclusions}
\label{sec:conclusions}

This work was a first step in evaluating the potential of mobile phones as stress detectors in working environments. The data was collected in an \emph{unconstrained} environment with \emph{unknown} stressors. We used accelerometer data to characterise subjects' behaviour by extracting time domain and frequency domain features. Then, statistical models were built to classify different self-reported stress levels. For our experiments, we also evaluated an ordinal classification method, which had no improvement in the overall performance, possibly due to the small number of classes (just 3). \emph{User-specific} models performed the best since they are targeted for each specific user but they require more labelled data. On the other hand, \emph{general} models had a lower overall performance but they don't require user specific labelled data which is sometimes tedious and time consuming to record. We proposed a \emph{similar-users} model in which a small amount of labelled data is used to find similar users and a classifier is built. According to our results, this proved to be a middle point between \emph{general} and \emph{user-specific} models, allowing a future system to begin providing feedback to the users on the onset, using \emph{general} model and as more labelled data is available \emph{similar-users} and \emph{user-specific} models could be built. The results we achieved are similar to the results found during literature review, with the difference that in our work we used a single accelerometer sensor only. This could open the possibility to implement a stress recognition system in personal fitness devices, which currently track physical activity only. Our follow-up study will extend data collection period to several months and include higher number of users in the experiments. Further analysis will focus on analysis of specific situations when the person is handling the phone (such as during phone call, text writing), which may provide a more fine grained insight into the users' behaviour.

%


\section*{Acknowledgements}
Enrique Garcia-Ceja would like to thank Consejo Nacional de Ciencia y Tecnolog\'{i}a (CONACYT) and the AAAmI research group at Tecnol\'{o}gico de Monterrey for the financial support in his PhD studies. 

\ifCLASSOPTIONcaptionsoff
  \newpage
\fi



\bibliographystyle{IEEEtran}
\bibliography{references}
\end{document}